\input epsf.tex
\documentstyle[12pt]{article}
\begin{document}
\begin{flushright}
LSUHE No. 343-2000
\end{flushright}
\def \beq{\begin{equation}}
\def \eeq{\end{equation}}
\def \ps{\psi}
\def \pb{\bar \psi}
\def \gi{\gamma_{i}}
\def \go{\gamma_{0}}
\def \g5{\gamma_{5}}
\baselineskip=24pt
\vspace{1.5cm}
\begin{center}
\bf{ Vortices in $SU(2)$ lattice gauge theory}
\end{center}
\vspace{1.0cm}
\noindent
\begin{center}
\rm{Srinath Cheluvaraja} \\
\it{Dept. of Physics and Astronomy, Louisiana State University,
Baton Rouge, LA, 70803} \\
\end{center}

\noindent{\bf{ABSTRACT}}\\
\rm
We investigate some properties of thick vortices and thick monopoles 
in the $SU(2)$ lattice gauge theory
by inserting operators which
create these excitations. Some quantities associated with the
dynamical behaviour of thick vortices and thick monopoles
are studied. We measure the
derivative of the free energy 
of the vortex
with respect to the coupling
and we find that it falls exponentially with the cross sectional
area of the vortex size.
We also study the monopole-monopole potential energy for thick
and thin monopoles.
Our results suggest that vortices and monopoles of increasing thickness
will play an important role in the large $\beta$ limit.

\vspace{0.5cm}
\begin{flushleft}
\end{flushleft}

\newpage 

The role of vortices in determining the properties of lattice gauge theories(LGTs)
has been under discussion for a long time. Vortices are topological
excitations which give rise to non-removable phase factors in the Wilson
loop. Vortices in non-abelian gauge theories \cite{hooft} can be defined using the
center of the group ($Z(N)$ for the group $SU(N)$)
-- thus defined, they are the natural generalizations of the
Peierls contours in the two-dimensional Ising model. Though these vortices are not
the most general vortices that can be defined in a non-abelian theory, they
can play an important role in the theory and their properties need to be
carefully studied. 
The analogy between
vortices and Peierls contours was used in \cite{mack} to arrive at certain
rigorous bounds for the Wilson loops in $SU(2)$ lattice gauge theories.
Vortices in lattice gauge theories were also studied in
\cite{yoneya}.
The role of vortices was further investigated in \cite{tombo} by rewriting $SU(2)$ LGTs
in terms of $SO(3)$ and $Z(2)$ variables. It was noted in \cite{mack} that 
the continous and non-abelian nature of the group $SU(2)$ can give rise 
to a different kind of vortex --a thick vortex -- whose core
has a thickness of more than one lattice spacing
. It was proposed \cite{mack,tombo} that these thick vortices can
have an important bearing on the confining
 properties of the theory. The thick
vortices have their analogues in the domain walls of ferromagnets with a
continuous symmetry--they are the generalizations of thick Peierls contours.
The free energy of these vortices can be made arbitarily small by
spreading them over large regions.
The thick vortices should be distinguished from the thin vortices --thin
vortices are like the vortices in a pure $Z(2)$ gauge theory--
their core has a thickness of only one lattice spacing.
It must be noted that these vortices are only one  of the many possible
excitations of the gauge theory. The vortex theory of confinement aims
to show that these vortices are sufficient, and perhaps necessary, to
produce the disordering of Wilson loops and
causes them to decay as the area of the minimal surface tiling the loop.
In this sense the vortices are expected to produce the same kind of
disordering of the Wilson loops as the Peierls contours do for the
spin-spin correlation function of the Ising model.
In order to quantify the vortex theory of confinement we have to
first identify the vortices, study their physical effects,
and then show that
the minima of the free energy favours
a condensate of these excitations. Addressing the complete
dynamical issue is quite complicated and in this paper we will present some
measurements which probe some properties of the thick vortices.
     
We should mention that there are different approaches to studying vortices
in non-abelian gauge theories which are being pursued at present.
One of them is the center projection approach.
In this approach \cite{proj1}, vortices are defined after
the theory is fixed to a particular gauge which still allows the vortex
degrees of freedom. This gauge, referred to as the maximal center gauge,
gets rid of many other degrees of freedom but retains the vortex excitations.
Vortices are defined in the gauge fixed theory just as in a $Z(2)$
LGT.
Studies of vortex
excitations in this gauge have led to the observation of center dominance--
the ungauged degrees (the center degrees of freedom) are sufficient
to reproduce the string tension and some other non-perturbative quantities.
More discussion of this approach can be found in \cite{proj2,proj3,proj4}.
Another approach developed further in \cite{kovtom} is to look at monopoles
and vortices using the
$Z(2)XSO(3)$ decomposition of the $SU(2)$ LGT. In this
approach the role of the Wilson loop as a vortex counter is quite
transparent and it is capable of addressing the issues of thin and thick
monopoles and vortices. Studies of vortices in this approach were
recently carried out in \cite{hay}.

We will first recapitulate what we mean by center vortices and how they arise in
lattice gauge theories. Much of this is well known and we repeat it merely
to set the definitions and to make our discussion self-contained.
We will talk about vortices
first in three, and then in four space-time (Euclidean) 
dimensions. In three dimensions a 
vortex is said to pierce a two dimensional region 
(simply connected) $R$ if every Wilson loop
surrounding this region picks up a phase ($Z(N)$ for the group $SU(N)$).
By definition, a vortex is a global disturbance which cannot be
confined into any finite region.
The effect of this vortex can also be understood as the action of
a pure gauge transormation
which is not single valued on every closed loop surrounding $R$. Because of
topological considerations the region $R$ cannot be shrunk 
to a point by
regular gauge transformations and this region is associated with the
core of the vortex . It is this region which carries the energy density of
the vortex. 
The vortex can extend in the dimensions orthogonal to $R$ and
can either stretch indefinitely, form closed loops, or end in objects( monopoles) which absorb the vortex flux.
In four dimensions the above picture gets repeated on every
slice in the extra dimension and the vortex line becomes a vortex sheet whose
area is the product of the length of the vortex line and the duration in time for
which the vortex propagates. These vortex sheets can 
either form closed two
dimensional surfaces or they can end in monopole loops.
The well known example of a vortex solution in a gauge
theory is the Nielsen-Olesen vortex which occurs in three-dimensional 
scalar QED \cite{niels} with a Higgs potential.
This vortex is the relativistic generalization of the
Abrikosov vortex seen in a type II superconductor.
Apart from the Nielsen-Olesen vortex there are not
many vortex solutions known, especially in unbroken non-abelian gauge theories.
However, the Nielsen-Olesen vortex carries an integral vortex charge
and is not a center vortex (which carries a $Z(N)$ charge).
Next we come to the definition of vortices on the lattice. First we will
consider a $Z(2)$ gauge theory whose action is given by
\beq
S=\beta\sum_{p}\sigma(p) 
\quad ,
\eeq
with the $\sigma(p)s$ given by $\sigma(p)=\prod_{l\in p}\sigma(l)$.
In a $Z(2)$
gauge theory a vortex configuration is a set of co-closed 
\footnote{ A set of plaquettes in a three(four) dimensional
lattice is said to be co-closed if
the links(plaquettes) dual to it form a closed loop(surface).}
plaquettes with $\sigma(p)=-1$. It follows from their definition that
Wilson loops linked by these configurations
will have a negative sign.
In three (four) space-time dimensions 
these vortices cost an action which is proportional to their
length (area), and they can condense in a phase in which the entropy
overwhelms their loop(surface) energy 
( i.e. when the free energy density of a vortex becomes
negative).
However, there is also a phase in which
the vortices are energetically suppressed (the free energy density of a vortex
becomes positive). The two phases of the $Z(2)$ LGT (in three and
four dimensions) can be distinguished by the behaviour of these
vortex excitations.
It is the aim of the vortex theory of confinement
to extend and adapt the above ideas to the $SU(2)$ (or $SU(N)$) LGTs.
The action for the $SU(2)$ LGT can be chosen to be the Wilson action
\beq
S=\frac{\beta}{2}\sum_{p}tr_{f}U(p)
\quad ,
\eeq
where $U(p)$ are the usual plaquette variables. It is possible to define a
vortex in the $SU(2)$ LGT just as in the $Z(2)$ LGT by using the sign of the
trace of the plaquette variable. A thin vortex is defined to be
a co-closed set of plaquettes
for which sign(trU(p))=-1. This definition of the vortex can always be made 
because just by restricting the $SU(2)$ elements to their center values
we should regain the vortex of the $Z(2)$ theory. However, the non-abelian
and continuous nature of the group $SU(2)$ allows us to define another
kind of vortex in the $SU(2)$ theory. This vortex, to be called a thick
vortex, has the property that sign(trU(p))=+1 for all the plaquettes,
but nevertheless Wilson loops pick up a center element whenever they
surround the vortex. It is obvious that such configurations can never appear
in the $Z(2)$ theory or, infact, in any abelian theory. In any abelian
theory the Wilson loop surrounding a region can always be decomposed into
the plaquettes that are tiling its minimal surface, and the Stokes theorem
ensures that if all the plaquettes are positive, every Wilson loop
is also positive. Classical vortex solutions with these properties have 
also been recently studied by the author \cite{myvort}.
An important
property of these vortices is that, because they they come with sign(trU(p))=+1,
they need not be suppressed in the $\beta 
\rightarrow \infty$ limit, unlike the thin vortices,
 and they can play an important role in the zero
lattice spacing limit. Apart from these excitations, there are other
excitations that can absorb the flux of thin and thick vortices. These
excitations will be referred to as $Z(2)$ monopoles following the
terminology in 
\cite{mack}. 
A thin vortex can
end in a thin $Z(2)$ monopole which (in three space-time dimensions) is defined on an elementary 3 dimensional 
cube $c$
violating the Bianchi identity. The thin $Z(2)$ monopole density 
on a cube is
given by
\beq
\rho_{1}(c)=(1/2)(1-sign(\prod_{p \in \partial c} trU(p)))
\quad .
\eeq
Analogously, a thick vortex can end in a thick $Z(2)$ monopole and  this
configuration violates the Bianchi identity on a 3 dimensional cube of side
$d$ lattice spacings . 
Its appropriate density (in three space-time dimensions) is given by
\beq
\rho_{d}(c)=(1/2)(1-sign(\prod_{d \in \partial c} trU(d)))
\quad ,
\eeq
where the product is taken over all $dXd$ plaquettes bordering a cube of
side $d$.
The subscript $d$ indicates that the density can be defined on any
3 dimensional cube of side $d$ lattice spacings.
In four space-time dimensions the monopoles become loops on the dual
lattice.
It is evident that both, thick vortices and
thick monopoles,
are possible
only in a non-abelian theory. Though the above excitations have only been 
defined on a lattice, it is a subtle question whether they will
go over to physical vortices in the continuum limit. According to the arguments
in \cite{hooft}, a scalar field in the adjoint representation can
dynamically arise in the gauge theory and  create these vortex
excitations. If this scalar field field develops a non-zero vacuum expectation
value, these vortices can condense and can cause quark
confinement. The very possibility that the thick vortices on the lattice
can go over to the dynamically generated vortices in the continuum
merits a further investigation of their properties.

In this paper we take a step towards studying the properties of these
excitations.
We introduce a term in the partition function which creates a stress
in the system and we study the effect of this stress on the system.
More specifically, we add a term to the Wilson action
\beq
S^{\prime}=\frac{\beta}{2}\sum_{p}trU(p) -\frac{\beta}{2}\sum_{p^{\prime}}trU(d)
\quad .
\eeq
The extra term is a Wilson loop variable defined over a square of side $d$ and 
it comes with
a negative sign. This term is introduced at a point in the (say) $ (12)$
plane, and it extends in the $3$ direction, and the term repeats
itself in the $4$ direction.
The effect of the extra term is to push the system such that
$trU(d)$ is negative, the effect of the usual Wilson action being to push
the system such that $trU(p)$ is positive. The above term
can also be understood as a way of implementing certain boundary conditions
for the system. By creating this stress we are probing the non-abelian
nature of the group in an essential way. In an abelian theory, the Stokes
theorem enables us to reduce any trace over a large loop to a product
of the traces over subloops, for instance
\beq
trU(C)=\prod_{p\in C}trU(p)
\quad .
\eeq
The term that we have introduced in the action explicitly violates the
above constraint.
We mention here that such stresses in gauge theories were introduced
in \cite{twist} and
have been studied before but all the studies
carried out so far
consider the special case in which only the plaquettes appear with an opposite
sign. Some recent works which discuss the case of the flipped plaquette are 
\cite{others}.
This corresponds to the case $d=1$ but we will be interested in the
case $d>1$ and how the stress behaves as a function of $d$.
We would like to measure the change in the free energy of the system before
and after applying this stress. From the previous discussion, the stress that
we are introducing into the system is nothing but a thick vortex of thickness
$d$ piercing a region in the $(12)$ plane and wrapping around the lattice
in $3$ direction. On the other hand, if the stress began at a point, say
$z=z1$, and stopped at, say $z=z2$, we would have a thick monopole-antimonopole pair bound
together by a thick vortex line. Since a $Z(2)$ antimonopole is the same as
a $Z(2)$ monopole we will always refer to the antimonopole as a monopole.
By introducing these stresses we can get some information about the
properties of the monopoles and vortices that we have discussed earlier.
Since direct free energy measurements are quite
forbidding we appeal to a simpler method in order to determine the affect of
this stress.
We first write
\beq
\mu_{d}=\frac{Z(\beta^{\prime})}{Z(\beta)}
\quad .
\eeq
$Z(\beta^{\prime})$ is the partition function of the system after applying the
stress and the $Z(\beta)$  is the partition function without the stress.
$\mu_{d}$ can also be written as
\beq
\mu_{d}=<\exp-\sum_{p^{\prime}}trU(d)>_{0}=\exp(-(F_{d}-F_{0}))
\quad ,
\eeq
where the subscript $0$ refers to the original partition function.
We have also expressed $\mu_{d}$ as the exponential of the free energy difference
between the stressed system and the system without the stress.
More precisely, $F_{d}-F_{0}$ measures the free energy difference when an
\it {additional} \rm thick vortex (having a thickness $d$)
is introduced into the original system (which may already contain thick vortices). Hence the effect of a large number of vortices already present is subsumed
in this free energy difference.
It is seen easily that
$\frac{\partial \log \mu_{d}}{\partial \beta}$ is
\beq
\frac{1}{\mu_{d}}
\frac{\partial \mu_{d}}{\partial \beta} =<\frac{1}{2} \sum_{p}trU(p) -
\frac{1}{2} \sum_{p^{\prime}} trU(d)>_{1}
-<\frac{1}{2}\sum_{p} trU(p)>_{0}
\quad .
\label{maineq}
\eeq
The subscript $1$ indicates that the average is taken with respect to the
stressed partition function, whereas the subscript $0$ indicates that the
average is taken with respect to the unstressed partition function. This
quantity directly measures the derivative of the free energy difference
because
\beq
\frac{\partial \log \mu_{d}}{\partial \beta}= \frac{-\partial(F_{d}-F_{0})}{
\partial \beta}
\quad .
\eeq
Before we present the numerical results of this measurement let us see how the
quantity looks in the strong coupling limit.
As mentioned before,
the quantity $\mu_{d}$ is also given by
\beq
<\exp -\beta \sum_{p^{\prime}}trU(d)>_{o}
\quad .
\label{noisy}
\eeq
Expanding the exponential to second order in $\beta$ we get
\beq
1-\beta \sum_{p^{\prime}}<trU(d)> + O(\beta^{2})
\quad .
\eeq
The second term is just a measure of the $dXd$ Wilson loop and goes
like the area law in the strong coupling limit, therefore we get
\beq
1-\beta (N_{3}N_{4})\exp (-\sigma d^2) +O(\beta^2)
\quad .
\eeq
($N_{3}$,$N_{4}$ are the sizes of the lattice in the $3$ and $4$ directions)
If the area of the $dXd$ plaquette is much larger than the area of the transverse
directions, the second term is quite small and after taking the logarithm
we get 
\beq
(F_{d}-F_{0})=\beta (N_{3}N_{4})\exp (-\sigma d^2)
\quad .
\eeq
The free energy per unit area of the vortex is given by
\beq
\sigma_{d}=\beta \exp(-\sigma d^2)
\quad .
\eeq
We thus find that in the strong coupling region the free energy per unit
area of a thick vortex decreases exponentially with the area of the 
vortex core. Note the relation between $\sigma_{d}$ ( free energy density of a
thick vortex of size d)
and $\sigma$ (the string tension). This calculation shows that
thicker vortices are  energetically more favourable
than thinner vortices.
The above calculation is valid only in the
strong coupling region. In addition, it required a delicate handling of
the area of the transverse directions as compared with the area of the
vortex. The limit $d\rightarrow \infty$ must be taken before the
limit $N_{3}N_{4}\rightarrow \infty$. We also note that the derivative of
the free energy of a thick vortex will also fall off exponentially with the
area of the vortex core. A direct measurement of the quantity in
Eq.~\ref{noisy} is also possible but since this quantity has large
fluctuations we have chosen to measure the quantity in Eq.~\ref{maineq}.
In a simulation we can measure
$\frac{\partial \log \mu}{\partial \beta}$
in the weak
coupling region and we will show that the exponential fall off with the area 
also persists in the weak coupling region.

In Fig.~\ref{2.3}, Fig.~\ref{2.4}, and Fig.~\ref{2.5} we present results at 
three different coupling
parameters for $\frac{\partial log \mu}{\partial \beta}$ as a function
of the area of the thickness of the vortex. In each case we find that there is an
exponential fall-off with respect to the area of the vortex. A logarithm of
the curve plotted vs the area of the vortex gives an approximately good
straight line fit (see Fig.~\ref{strfit}). The simulations were done on a lattice whose
spatial extent was $10X10$ and the remaining size was $6X6$. Vortices
of thickness ranging from $1$ to $5$ were studied, and in
order to avoid finite size effects associated with these vortices
the lattices were chosen to have a spatial extent of atleast twice the
largest vortex size. The other parameters of the lattice were fixed by
some limitations in the computer time. It must be mentioned that each
point in Figs.1-3 requires a separate simulation because
Eq.~\ref{maineq} is to be used for different values of $d$.
300,000 data points
were gathered at each point and the errors were estimated by binning.

We wish to point out that though our result is for
the derivative of the free energy of the vortex area as a function of the
coupling, the free energy of the vortex will also fall-off exponentially with
the area, just as in the strong coupling limit.
This result has two immediate implications. 
It indicates that as we approach
the weak coupling limit ($\beta \rightarrow \infty$), vortices of 
larger and larger area are more favourable than vortices of any fixed
area. By spreading the core of the vortex over a large area the
energy of the vortex loops can be made arbitrarily (exponentially) small. 
This feature
is quite like the thick Peierls contours of ferromagnets with a
continuous symmetry.
It also means that in the region of the continuum limit we must
be able to tackle a many body vortex problem in
which vortices can have large overlaps with each other. This feature seems to
make the study of such vortex gases quite difficult. However,
from the point of view of the continuum limit, very large vortices are
almost necessary;
if lattice vortices are
to survive as continuum excitations they must appear with a length scale
which diverges in lattice units  as the lattice spacing goes to zero. Only then can they
yield a vortex corresponding to 
some physical thickness. Thick vortices indeed admit
this possibility by appearing with arbitrarily large thicknesses.

We now turn to thick monopoles. As mentioned earlier, if the stress is
taken to extend only between two points, say $z=z1$ and $z=z2$, it corresponds
to an external thick monopole-monopole loop running along the $4$ direction
at $z=z1$ and $z=z2$ and separated by a thick vortex sheet between $z=z1$ and
$z=z2$. We again consider Eq.~\ref{maineq} but this time since the
stress is defined only between two points we now write
\beq
\mu_{d}(x,y)=\frac{Z(\beta^{\prime})}{Z(\beta)}
\quad .
\eeq
This quantity can now be written as
\beq
\mu_{d}(x,y)=<\exp-\sum_{p^{\prime}}trU(d)>_{0}=\exp(-(F_{d}(x,y)-F_{0}))
\quad .
\eeq
Eq.~\ref{maineq} is now the quantity
\beq
\frac{-\partial(F_{d}(x,y)-F_{0})}{
\partial \beta}
\quad ,
\eeq
and it measures the derivative of the potential energy with respect to the
coupling of 
an external
monopole-monopole pair as a function of their separation.

It is interesting to see how this mutual energy depends
on their separation.
We expect that if thick monopoles abound
in the system there will be a screening mechanism in operation which
screens the interactions between an externally introduced monopole-monopole
pair.
On the other
hand, if the density of monopoles is very low then the screening mechanism
is no longer operative. Before we study their interaction energy we
will first say a few things about thick and thin monopoles.
We stress again that the thick monopoles
and thick vortices are not suppressed by the term in the 
action as they are configurations
where $sign(tr(U(p))=+1$ everywhere. It is well known that thin $Z(2)$
monopoles are suppressed at weak coupling \cite{mack,tombo,brow}. The thick
monopoles need not be (and are not) similarly suppressed. Just for
comparison we show the behaviour of thick vs thin monopoles as a function
of the coupling in Fig.~\ref{thickvsthin}. We see that thick monopoles 
continue to be dense even after all the thin monopoles have disappeared.
However, thick monopoles \it{also} \rm start disappearing as we go to weaker and
weaker coupling, although much slowly. To see the fall in the density of thick monopoles we
have to go to larger and larger values of $\beta$. Fig.~\ref{thickvsthin}
shows that as we go to larger values of $\beta$ even thick monopoles start falling off
but monopoles of even larger thicknesses are still abundant. 
There is no obvious argument (like there is for thin monopoles) 
for the suppression of thick monopoles with
increasing $\beta$. The behaviour of thick monopoles seems to be more subtle.
Similar considerations apply to thick vortices.
We can also measure
the density of thick vortices piercing a loop $C$ by looking at the
instances of the negative sign for the following quantity
\beq
\ tr\  U(C) 
\prod_{p \in C}\theta(\eta(p))
\quad .
\eeq
 $\eta(p)$ is the sign of the trace of the
plaquette variable $tr U(p)$. A quantity $\rho_{v}(C)$ can be defined to take
the value $1$ whenever the above quantity is negative and $0$ otherwise.
The $\theta$ function ensures that
all the plaquettes inside the loop have a positive trace. A negative 
value for the above quantity 
implies a thick vortex piercing the 
minimal surface (in this case the plane containing the loop)
bounding the loop. 
Since a direct measurement of the vortex density is
complicated by the fact that vortices can bend and twist in every possible
way, we have chosen to measure the density of vortices piercing a unit
area in the $1-2$ plane. Though this quantity does not give complete information about the
structure of the vortices, for example it does not give any information
about the size distributions of the vortices, it does indicate the number of vortex lines
piercing a plane, the high density of which is atleast a necessary
condition for any vortex theory of confinement.
The sign of this quantity is measured for
different planar (in the 1-2 plane at z=t=1) loops that can be drawn at a point and then the average is taken
over all the points in the plane. 
We have
looked at square loops having sizes of up to $d=6$. Fig.~\ref{vordens} shows this
density at different values of the coupling. 
This plot shows that, just like the
monopoles, the vortex density also decreases with increasing $\beta$ but
then vortices of larger thicknesses are still significant.
The data in Fig.~\ref{thickvsthin}
and Fig.~\ref{vordens} were generated on a $8^{4}$ lattice after collecting
10000 data points.

We now present some measurements of the monopole-monopole potential energy
as a function of their separation distance for monopoles of different
thicknesses. Fig.~\ref{monpotw=1} shows the potential energy for
monopoles of thickness $d=1$ at $\beta =1.0$ and $\beta=2.4$.
The approximately linear rise at $\beta=2.4$ shows that
their separation energy directly increases with the separation distance.
By contrast the curve for $\beta=1.0$ is quite flat.
At $\beta=2.4$, the thin monopoles have become very rare and the screening
mechanism between an external monopole-monopole pair is absent and their
interaction energy is directly proportional to the length of vortex
line joining them. On the other hand, at $\beta=1.0$, when the monopole density is quite
large, the energy of the external pair does not increase linearly with
their separation because of the screening effect of the 
large number of monopoles in the system.
A screening by the monopoles in the system should produce
a Yukawa like interaction between the external
charges. A similar measurement for
thick monopoles shows very different behaviour.
Fig.~\ref{monpotw=3}
shows the monopole-monopole potential energy for a thick monopole
of thickness $d=3$.
At $\beta=2.4$ the thick
monopole-monopole potential energy curve is quite flat. This can again
be understood from the presence of a large number of thick monopoles ($d=3$)
in the system. See Fig.~\ref{thickvsthin}. At larger values of
$\beta$, for example $\beta=4.0$, the thick monopoles appear to show a
screening behaviour. The main difference between Fig.~\ref{monpotw=1}
and Fig.~\ref{monpotw=3} is the signs of the slopes of the curves in the
two plots. The slope of the curve is the negative of the 
derivative of the free energy of the monopole pair with respect to the
coupling.
While the first plot (with negative slope) shows a linear rise 
with the separation distance in the interaction
energy of a thin monopole-monopole pair at $\beta=2.4$, the second plot ($\beta=4.0$) 
shows a decrease (with positive slope) in the interaction. 
This different behaviour for monopoles of finite
thickness compared with monopoles of unit thickness again shows that the self-
energy of thick monopoles has a more subtle behaviour than that of the
thin monopoles. A possible explanation of this difference in behaviour is that
thick monopoles tend to repel each other (like similarly charged particles)
and lower their potential energy. The data in Fig.~\ref{monpotw=1} and
Fig.~\ref{monpotw=3} were taken on a $6^{4}$ lattice after 50000 iterations.

We wish to put our observations in the context of, what has been called, the
hierarchichal
$Z(2)$ theory of confinement \cite{effz2,brow}. 
A hierarchy of effective $Z(2)$
theories is supposed to be operating at larger and larger values of
$\beta$. Each member of this hierarchy is an effective $Z(2)$ theory
defined over a different scale. As the weak coupling limit is approached
the dynamics can be effectively described by a different effective
$Z(2)$ theory at larger and larger scales ( we ascend the hierarchy of effective
$Z(2)$ theories). For every $\beta$ there is an associated scale $d$ 
(the square root of the time constant of the exponential in Figs.1-3
) at which
there is an effective $Z(2)$ like behaviour.
As $\beta$ increases, $d$ also
increases.
This is one way in which the $SU(2)$ LGT
can escape the phase transition seen in the $Z(2)$ LGT. It is not easy to
write down the effective $Z(2)$ theories for the different members of this
hierarchy. In \cite{brow} an effective $Z(2)$ theory
is derived for the $d=1$ case in
the strong coupling limit. Some properties of the effective
$Z(2)$ theory are discussed in \cite{effz2}. 
The approach presented in this paper
of directly measuring some properties of thick vortices
may be useful in elucidating this hierarchy of $Z(2)$ theories.
We have also observed that as $\beta$ increases thicker monopoles and vortices
cannot be ignored, for every value of $\beta$ in the weak
coupling regime there seems to be a minimum thickness of the monopoles and
vortices although there is no maximum thickness.

The aim of this investigation was mainly to study the behaviour of
the additional extended objects, thick vortices and thick monopoles, which
the continuous and non-abelian nature of the group $SU(2)$ admits. The
behaviour of these excitations is quite different from the usual thin
vortices and thin monopoles that one is familiar with. Any vortex theory of
confinement must be able to handle these excitations in some, perhaps approximate,
way. The method presented in this paper can serve as a useful technique for
carrying out further studies of these objects. For instance, we can study the
second derivative of the free energy with respect to the coupling and look
for peaks to see the different excitations decoupling as we approach the
large $\beta$ limit.

\noindent
Acknowledgement: This work was supported in part by United States
Department of Energy grant DE-FG 05-91 ER 40617.

\newpage
\begin{figure}
\caption{ $\frac{\partial log \mu}{\partial \beta}$ vs $A$ at $\beta=2.3$.}
\label{2.3}
\end{figure}
\begin{figure}
\caption{ $\frac{\partial log \mu}{\partial \beta}$ vs $A$ at $\beta=2.4$.}
\label{2.4}
\end{figure}
\begin{figure}
\caption{ $\frac{\partial log \mu}{\partial \beta}$ vs $A$ at $\beta=2.5$.}
\label{2.5}
\end{figure}
\begin{figure}
\caption{ Fitting the exponential in Fig.~\ref{2.5} to a straight line.}
\label{strfit}
\end{figure}
\begin{figure}
\caption{ Comparison of thick vs thin monopoles as a function of $\beta$.}
\label{thickvsthin}
\end{figure}
\begin{figure}
\caption{$\rho_{v}(C)$ as a function of $\beta$ for loops of sizes 2-6.}
\label{vordens}
\end{figure}
\begin{figure}
\caption{ Potential energy of thin $d=1$ monopoles as a function of
distance at $\beta= 1.0$ and $\beta=2.3$.}
\label{monpotw=1}
\end{figure}
\begin{figure}
\caption{ Potential energy of thick $d=3$ monopoles as a function of
distance at $\beta=2.4$ and $\beta=4.0$.}
\label{monpotw=3}
\end{figure}
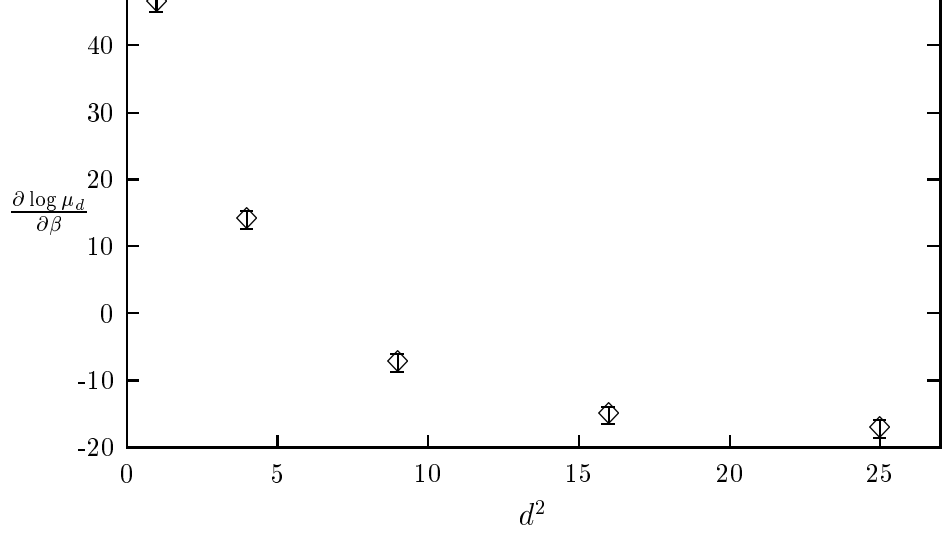
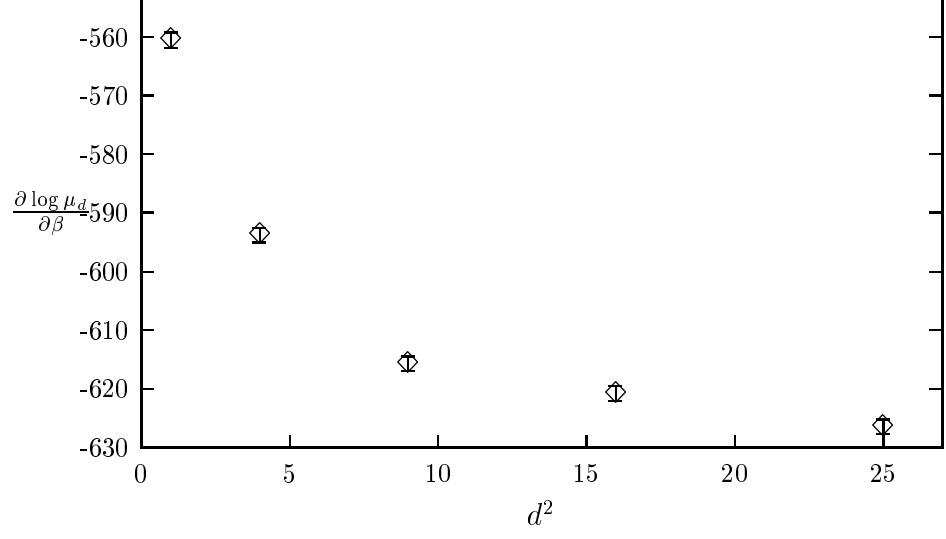
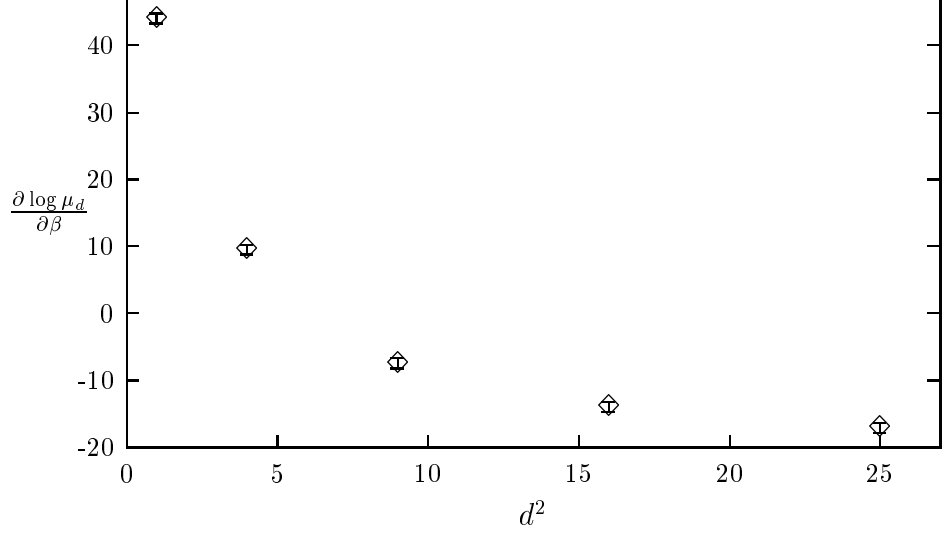
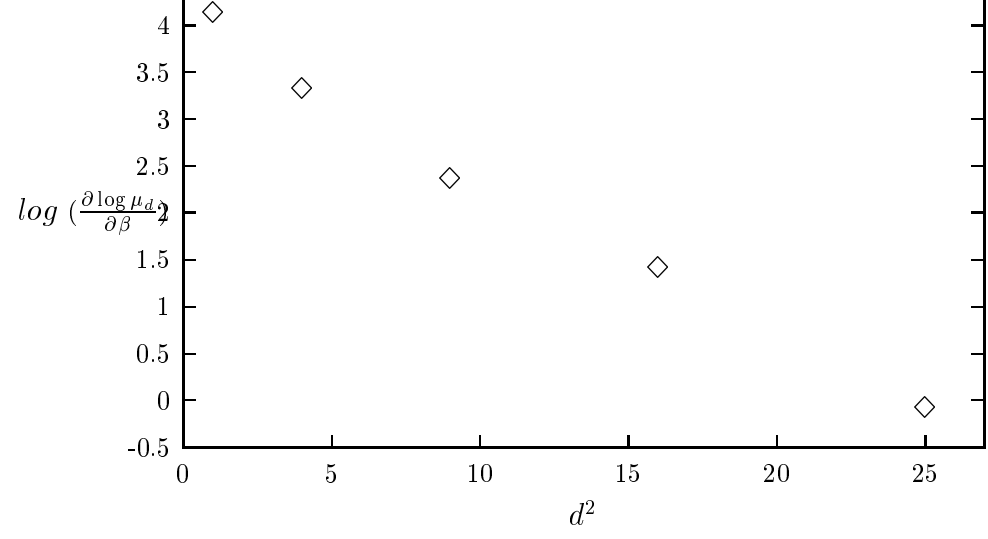
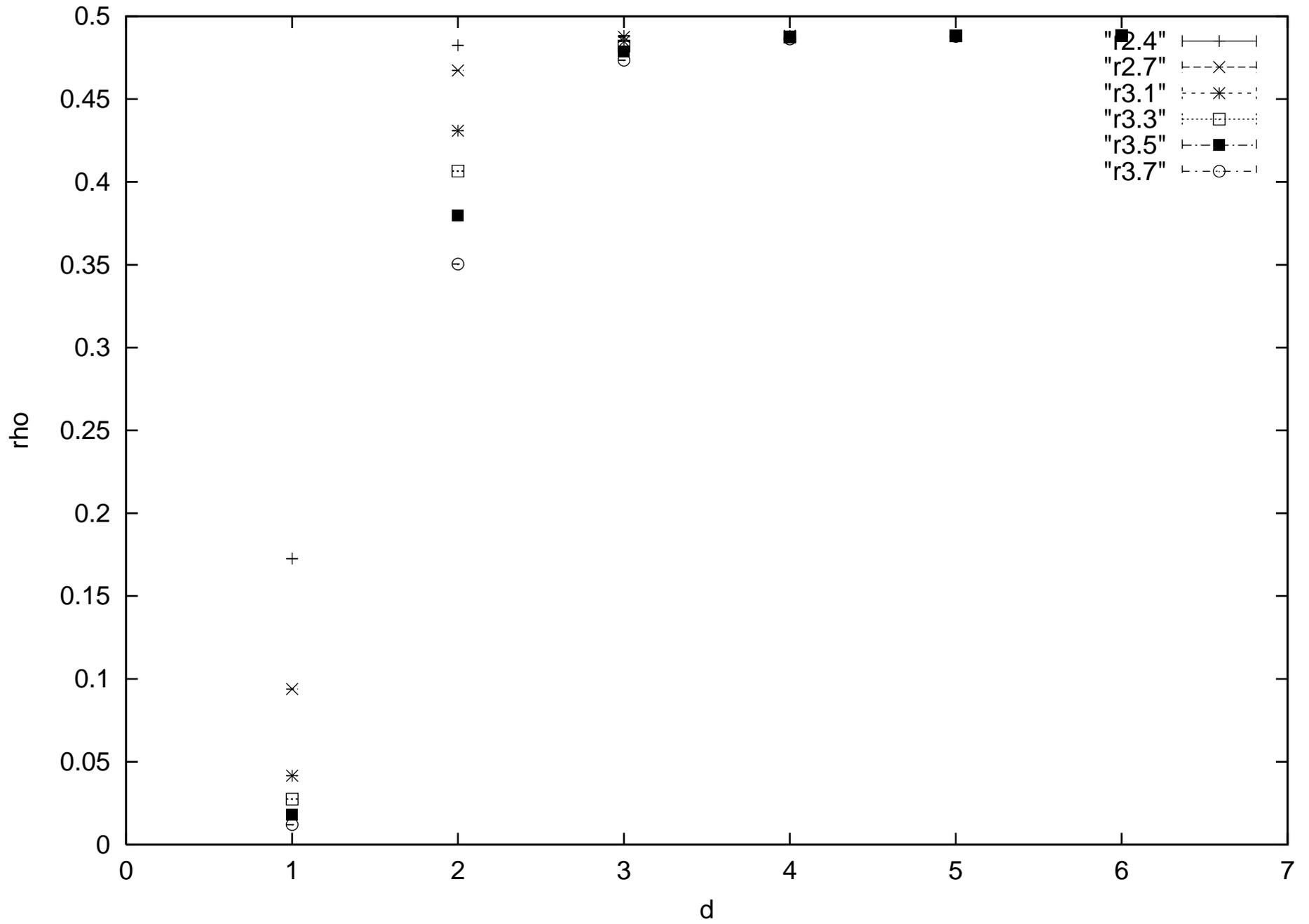
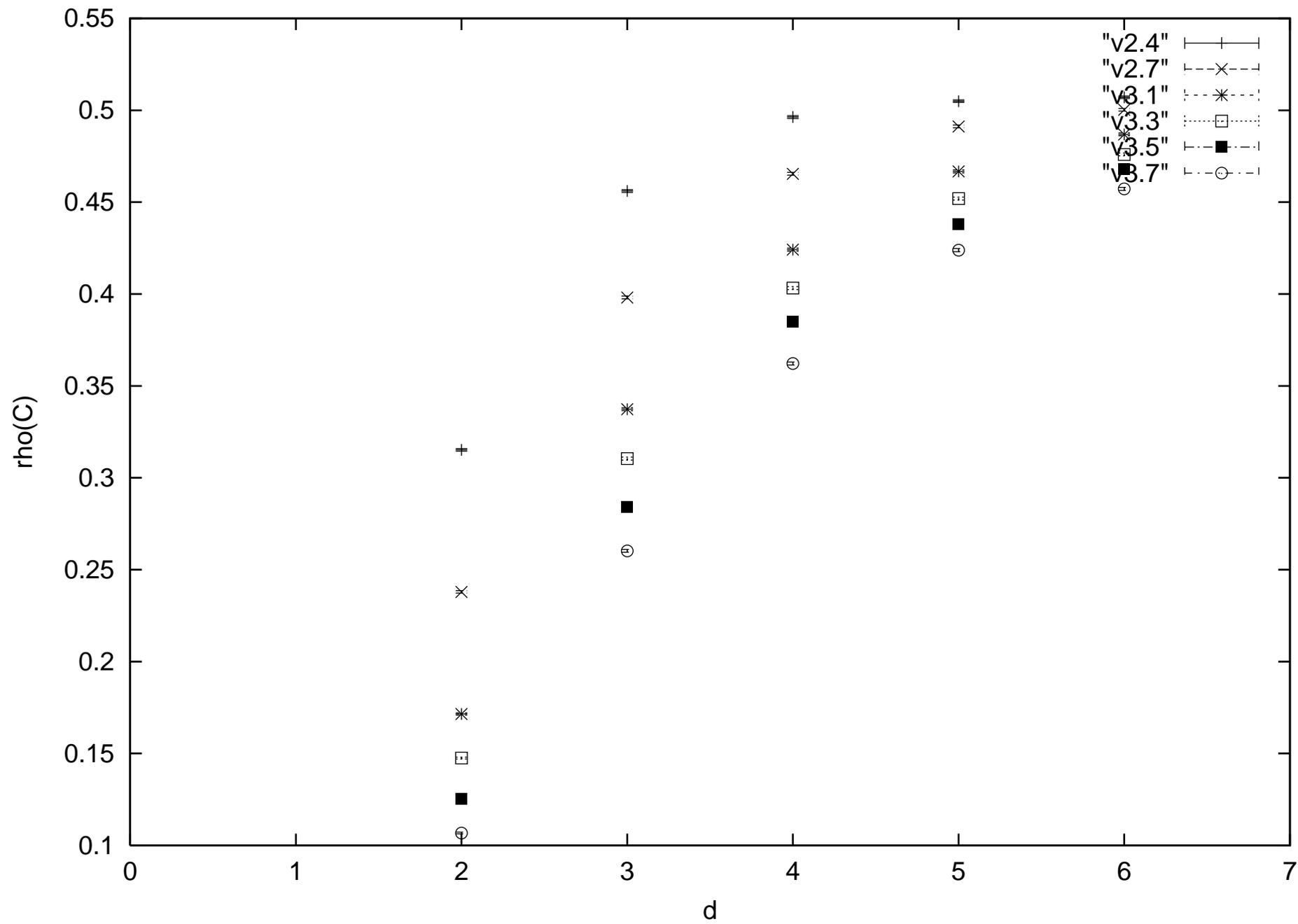
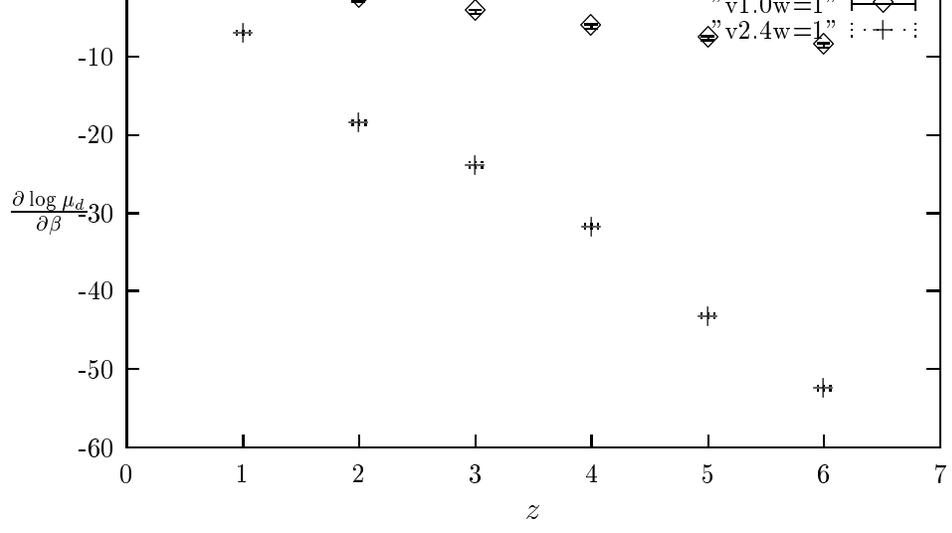
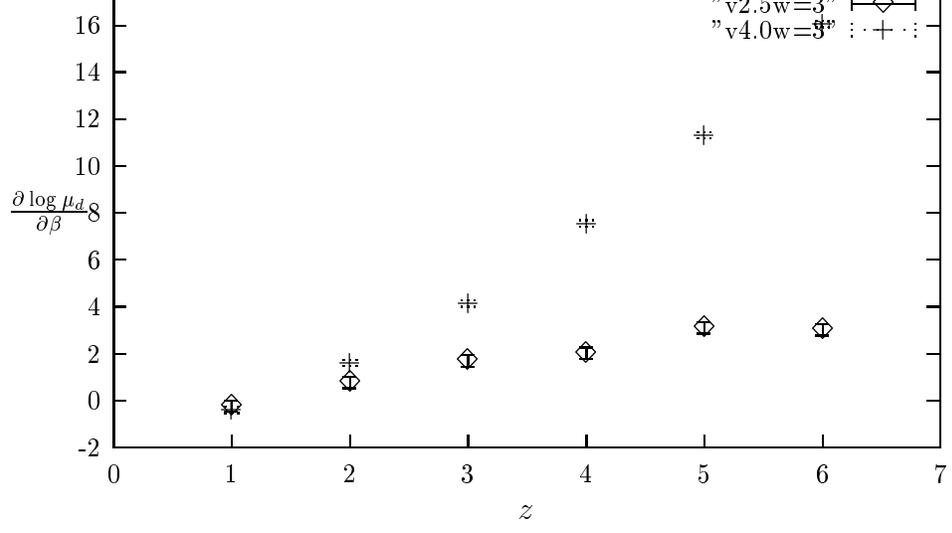
\end{document}